# Revisiting Dirac and Schrödinger: A Proof Offered for the Non-Relativistic Time-Dependent Schrödinger Equation

Ali Sanayei

*Institute for Theoretical Physics, University of Tübingen, Auf der Morgenstelle 14, 72076 Tübingen, Germany*

sanayeiali8@gmail.com

**Abstract**

Three formal pictures of quantum mechanics exist so far which can be transformed into each other by known techniques. In the Schrödinger picture, the fundamental pivot is a postulated general equation of motion for a state whose representation in the function space is the well-known non-relativistic time-dependent Schrödinger wave equation. There have been also carried out several attempts to either exemplify or derive the equation, however, most of them are not completely satisfactory. In this paper, three plausible axioms – which are acceptable for almost all physicists and also are not in contradiction with any other valid propositions in quantum mechanics – together with two definitions are employed to build an axiomatic framework, and then with the help of the Dirac formalism, it is demonstrated that the time-dependent Schrödinger wave equation is no longer a postulate (axiom or conjecture) for the whole theory, but a theorem. Subsequently, a proof for the theorem is presented. The result implies that the other remaining axioms of the picture essentially involve the Schrödinger equation, and this consequence lets the whole theory become stronger, because one of its elementary axioms has now become a theorem which has been derived from the others.

## 1. Introduction

It is customary that quantum mechanics has been formulated in three ways, by Heisenberg, Schrödinger and Feynman, respectively. In the Heisenberg picture, Hilbert-space operators are substituted for variables and equations of motion are written using commutators [2,3,5]. In the Dirac formalism, states correspond to fixed vectors, and dynamical variables correspond to moving linear operators [23,24]. On the other hand, in the Schrödinger picture, states correspond to moving vectors, and dynamical variables correspond to fixed linear operators [23,24]. The pivotal point in this picture is a general equation of motion for a state whose representation in the function space is the well-known time-dependent Schrödinger wave equation. The way that how exactly Schrödinger discovered the equation, is approximately clear. One of the most important historical documents we have on hand is an article by Felix Bloch [34], who was in the audience in the colloquia held at ETH, in one of which Schrödinger got motivated to search for the equation. Bloch mentioned that during one of those colloquia Schrödinger was stimulated by

Debye to work on the de Broglie's thesis. In one of the next colloquia Schrödinger gave a beautiful talk on that thesis, however, Debye was not totally convinced and casually remarked that, to deal properly with waves, one has to have a wave equation! That striking remark motivated Schrödinger to find an equation, and subsequently in a few weeks he gave another talk, and interestingly declared that he has found such an equation. Schrödinger also explained there what he was going to publish as a paper on the quantization as an eigenvalue problem, which was later called by Born 'the second dramatic surprise' [22].

As Bloch has stated, one thing is manifestly clear, that Schrödinger was motivated to work on the de Broglie's wave equation, which referred to force-free motion, and then he applied it to the case where the effect of force is taken into account. In this context, he started by an example of a mechanical system which can be formulated by the well-known Hamiltonian function of action, say $W$. Then he rewrote it in a way to conclude that any function $W$ of space alone can be described by giving geometrically the system of surfaces on which $W$ is constant. Next, he associated with those surfaces the idea of stationary sinusoidal waves whose phases are given by the quantity $W$. So he could write the wave function as a sinusoidal function whose amplitude is generally a function of positions, e.g. $A\sin(W/K)$, and concluded that the constant $K$ should have the physical dimension of action, and accordingly justified that it should be a universal constant. Via this way, the Planck's constant appeared in the equation. By writing the whole equation as a partial differential equation, he employed the quantum discontinuity condition of energy, and applied the equation to the hydrogen atom. Meanwhile, he found out that one of the results, which had been inferred from the discontinuity condition of the energy levels, corresponded exactly to the Bohr's stationary energy levels of the elliptic orbits. Ultimately, he generalized the whole problem to three dimensions and discussed the solutions [6,7,9,36]. Afterwards, the equation was received as one of the elegant postulates of quantum mechanics which can be applicable to any general problem. That's why for some time Schrödinger, himself, was not sure about the equation in a general respect [34]. Nevertheless he could present a proof on the equivalency of his picture with the Heisenberg's [8], whose great primary part was only guess-work, as Born later mentioned during his Nobel Lecture [22].

It is also worth mentioning that, in the meantime the whole theory had been reformulated by Dirac using a new formalism as well as by finding a relationship between commutators and Poisson brackets [4,24], which was later called by Born 'the first dramatic surprise' [22]. Dirac and Jordan, independently of one another, showed that it is possible to change the representation and move from the Heisenberg picture to Schrödinger's using the transformation theory [10,14,24,38,57]. In addition, Dirac published some papers on remarkable analogies between classical mechanics and the new quantum version by relying on the Lagrangian and the action principle, e.g. [13,16]. He also later emphasized several times this viewpoint during his lectures [27]. Dirac's outlook became a fruitful motivation for Feynman to introduce the third picture. He emphasized that this formulation of non-relativistic quantum theory was suggested by some remarks of Dirac concerning the relation of classical action to quantum mechanics [17,50]. In the third picture, the transitions are given by a set of trajectories in configuration or phase space, each contributing to the phase of a Hilbert vector, whose squared amplitude is the desired probability. In other words, the probability amplitude is associated with an entire motion of a particle as a function of time, rather than simply with a position of the particle at a particular time [17,50]. Interestingly, the mentioned third picture was originally ignited by Wentzel, whose paper [1] was published one year before the first Heisenberg contribution [40,41]. This point was also implicitly expressed by Bloch in his article of 1976 [34]. Therefore the third picture is in fact the first one which had been fallen into oblivion. Accordingly, one can call the first one ("third") on the 'WDF picture' whereby the letters WDF stand for three pioneer names: Wentzel, Dirac and Feynman.

In the Schrödinger picture, the most central part is his well-known equation which has since successfully been generalized as applying to all non-relativistic problems. It is often considered as a fundamental and elegant postulate. So when one looks at the quantum mechanics from the point of view of the Schrödinger picture, the equation is taken as the cornerstone. Starting from it, one can find different ways to go to other pictures and come back. It is noteworthy that, there have been several attempts to derive the Schrödinger equation from the first principles, see [28,45,56] and references therein. Most were motivated by the primary wave viewpoint in order to find a way to generalize the equation to the whole theory of quantum mechanics. The reason why they are unsatisfactory can be stated as follows: first, one cannot find in these approaches an exact formalism within which the derivations are performed. Some expressions remain vague or there is no recognizable basic framework used. In particular, the exact meaning of the term 'wave function' remains unaddressed. Dirac once remarked, although the 'wave function' was a reasonable name in the early days of quantum mechanics (as all the examples of these functions were of the form of waves), it is not a descriptive one from the point of view of the modern general theory, see p. 80 of Ref. [24]. The second reason for the lack of success is a problem coming from the use of the word 'wave equation.' Misled by this word, many authors have attempted to deal with the classical-mechanical wave equation and some particular waves in it, like plane waves, to derive a partial differential equation in which there exists a first-order time derivative and a second-order space derivative. That is why Julian Schwinger once noted that, he never believed that this simple wave approach was acceptable as a general basis for the whole subject, see p. 29 of Ref. [43]. And the third reason, quit a number of pertinent approaches have considered many postulates to derive or explain the equation. As long as the number of postulates increases, the strength and reliability of the theory proportionally decreases. Meanwhile, in spite of unconvincing efforts, there are a number of works that have tried to put forward the discussion in exact formalisms. In Section 4 we will present a technical discussion on them and will demonstrate that why our approach is distinct.

Now let us propound a question: is it perhaps possible to consider a few primary assumptions and make a formal axiomatic framework in which the time-dependent Schrödinger equation becomes a *theorem* rather than a *postulate*? If such a framework can be built, then the number of those axioms should not exceed the number of postulates which currently exist in the Schrödinger picture. In particular, they must be intuitively plausible and must not contradict other valid propositions in quantum mechanics. It may be asked whether such an approach is needed if we already have three pictures and one knows how to proceed from one to the other. The answer is that: if those given axioms are convincing enough – in the sense of being palpable, plausible and trivial for physicists – and then if the time-dependent Schrödinger equation is *derived* from those axioms, then the first advantage would be that, the whole theory is more powerful as at least one postulate has been eliminated. A second advantage is that, by accepting those remaining axioms, one can be sure that the validity of the equation in the given framework is not pending as long as a refutable experiment has not been yet found, because one axiom has now been replaced by a proof.

It may be notified that one can choose another picture of quantum mechanics, e.g. WDF, and accept its postulates to find an exact derivation of the time-dependent Schrödinger equation in the whole theory (as Feynman has beautifully shown [17,50]). Nevertheless, is it possible to derive the equation in the Schrödinger picture, *itself*? We claim that the answer is in the positive. Accordingly, the main aim of this note – motivated by the Dirac formalism (e.g. [24]) – is: (i) to build an axiomatic framework whose axioms are acceptable for almost all physicists and are fewer than the number of the postulates existing in the current Schrödinger picture of quantum mechanics; (ii) to consider the non-relativistic time-dependent Schrödinger equation as a *theorem* (rather than a postulate), and to present a proof within the given framework, that is, it will be demonstrated that the non-relativistic time-dependent Schrödinger equation can be inferred from the remaining axioms.

The paper is organized as follows: in the Section, called preliminaries, two definitions are presented. We employ only three axioms, namely, the spatial coordinates (positions) are observables and possess continuous ranges of eigenvalues, the total energy of a dynamical system (its Hamiltonian) is always an observable, and the superposition relationship between states remains invariant under displacements. Then the fundamental theorem is presented. Section 3, which is the main body of the paper, presents the proof of the theorem based on those three axioms. In this regard, some lemmas with their proofs are included. Section 4 is dedicated to a technical discussion about those works which have tried to present precise formalisms to deal with the problem. In that section we will show what the privilege of our approach is. Finally, the concluding remarks are presented in Section 5. An Appendix at the end explains a special point in one of the axioms of the framework.

## 2. Preliminaries

Throughout this paper, we consider a bra or a ket vector (or rather its direction) corresponds to a state of a dynamical system at a particular time, and the linear operators correspond to the dynamical variables at that time [24]. Planck's constant is considered as a universal constant.

**Definition 1.** An *observable* is a real dynamical variable whose eigenstates form a complete set [24].

**Definition 2.** In a representation in which the complete set of commuting observables $\xi_1,...,\xi_u$ are diagonal, any ket $|P\rangle$ will have a representative, $\langle\xi'_1...\xi'_u|P\rangle$, where $\langle\xi'_1...\xi'_u|$ denotes a general basic bra, and for brevity $\langle\xi'|P\rangle$. This representation is a definite function of the variables $\xi'$ denoted by $\psi(\xi')$. The ket is then simply written as $\psi(\xi)$, a function of the observables $\xi$. We call such a function the *Dirac function* [24].

**Remark 1.** Throughout this paper, we use the representation introduced in Definition 2.

**Axiom 1.** The spatial coordinates (positions) are all observables and have continuous ranges of eigenvalues.

**Axiom 2.** The total energy (Hamiltonian) of a dynamical system is always an observable.

**Remark 2.** How does the concept of "Hamiltonian" appear in this framework? Axiom 2 only expresses that the Hamiltonian is always an observable, however, it does not state what the meaning of Hamiltonian in this framework is. This subtle point is discussed in Appendix showing that the concept of Hamiltonian is meaningful in our framework.

**Axiom 3.** The superposition relationships between states remain invariant under displacements.

**Remark 3.** The exact meaning of Axiom 3 is that: if one takes a superposition relation, which holds for certain states at time $t_0$ and gives to a linear equation between the corresponding kets (e.g. $|Rt_0\rangle = a|Xt_0\rangle + b|Yt_0\rangle$), then the same superposition relation holds between the states of motion throughout the time during which the system is undisturbed, and therefore it will lead to the same equation between the kets corresponded to these states at any time $t$ (e.g. $|Rt\rangle = a|Xt\rangle + b|Yt\rangle$), provided that the system remains undisturbed in that time interval and also the arbitrary numerical factors by which those kets may be multiplied are suitably chosen (see Ref. [24] and p. 97 of Ref. [52]).

**Lemma 1.** Suppose that $u$ and $v$ being regarded as arbitrary functions of a set of canonical coordinates and linear momenta $q_r$ and $p_r$, respectively. If the Poisson bracket of $u$ and $v$ is denoted by $\{u,v\}$ and their commutation by $[u,v]$, then the following relationship holds between their Poisson bracket and commutation:

$$uv - vu \equiv [u,v] = i\hbar\{u,v\}, \qquad (1)$$

where $\hbar$ denotes the Planck's constant, $h$, over $2\pi$.

*Proof.*

By the definition of the Poisson bracket, e.g. [19], one can write:

$$\{u_1 u_2, v\} = \sum_r \left( \left( \frac{\partial u_1}{\partial q_r} u_2 + u_1 \frac{\partial u_2}{\partial q_r} \right) \frac{\partial v}{\partial p_r} - \left( \frac{\partial u_1}{\partial p_r} u_2 + u_1 \frac{\partial u_2}{\partial p_r} \right) \frac{\partial v}{\partial q_r} \right) = \{u_1, v\} u_2 + u_1 \{u_2, v\} \quad (2)$$

By analogous similar calculation, one obtains

$$\{u, v_1 v_2\} = \{u, v_1\} v_2 + v_1 \{u, v_2\}. \quad (3)$$

Using (2) and then (3) one can calculate the Poisson bracket $\{u_1 u_2, v_1 v_2\}$ as follows:

$$\{u_1 u_2, v_1 v_2\} = \{u_1, v_1 v_2\} u_2 + u_1 \{u_2, v_1 v_2\} = (\{u_1, v_1\} v_2 + v_1 \{u_1, v_2\}) u_2 + u_1 (\{u_2, v_1\} v_2 + v_1 \{u_2, v_2\})$$

$$= \{u_1, v_1\} v_2 u_2 + v_1 \{u_1, v_2\} u_2 + u_1 \{u_2, v_1\} v_2 + u_1 v_1 \{u_2, v_2\} \quad (4)$$

It is also feasible to calculate the same Poisson bracket using (3) and then (2), that is,

$$\{u_1 u_2, v_1 v_2\} = \{u_1 u_2, v_1\} v_2 + v_1 \{u_1 u_2, v_2\} = (\{u_1, v_1\} u_2 + u_1 \{u_2, v_1\}) v_2 + v_1 (\{u_1, v_2\} u_2 + u_1 \{u_2, v_2\})$$

$$= \{u_1, v_1\} u_2 v_2 + u_1 \{u_2, v_1\} v_2 + v_1 \{u_1, v_2\} u_2 + v_1 u_1 \{u_2, v_2\} \quad (5)$$

The right-hand side of (4) and (5) must be equal. Hence

$$\{u_1, v_1\}(u_2 v_2 - v_2 u_2) = (u_1 v_1 - v_1 u_1)\{u_2, v_2\}. \quad (6)$$

So (6) leads one to the following general conclusion:

$$u_1 v_1 - v_1 u_1 = \Omega \{u_1, v_1\}, \quad (7)$$

$$u_2 v_2 - v_2 u_2 = \Omega \{u_2, v_2\}, \quad (8)$$

where $\Omega$ does not depend on $u_i$ and $v_i$, and so it is a constant number that we shall find it [59]. It is evident that, $\Omega = 0$ is a trivial solution of (6) and then will show that $u_i$ and $v_i$ are ordinary variables. To find the general solution, one should find $\Omega$ such that the right-hand sides of (7) and (8) will not vanish. We also want the Poisson bracket of two real variables to be real. In this regard, suppose that $u_i$ and $v_i$ are real dynamical variables. Accordingly they are equal to their complex conjugates. Moreover, the Poisson bracket of those complex conjugate variables is the same as of the original variables. By making complex conjugate from both sides of (7) and noting that they have been already assumed to be real variables, one obtains

$$v_1 u_1 - u_1 v_1 = \overline{\Omega} \{u_1, v_1\}, \quad (9)$$

where $\overline{\Omega}$ denotes the complex conjugate of $\Omega$.

By comparing (7) and (9), one simply finds out that:

$$\Omega\{u_1, v_1\} = -\overline{\Omega}\{u_1, v_1\} \Rightarrow \Omega = -\overline{\Omega} \qquad (10)$$

Therefore the constant $\Omega$ should be a pure imaginary number. In addition, since $u_i$ and $v_i$ are both arbitrary functions of canonical coordinates and linear momenta, one can obtain that the physical dimension of the left-hand side of (7) or (9) in the international system of units is $m^2 \, kg/sec$, however the right-hand side is dimensionless. Since $u_i$ and $v_i$ are assumed to be general functions of canonical coordinates and linear momenta, we need a universal constant with the same physical dimension. The only universal constant with that dimension – which is known so far – is the Planck's constant. Thus: $\Omega = i\hbar$. The same discussion can be put forward to (8) and one would be able to arrive to the same result (Q.E.D).

**Remark 4.** From now on, Eq. (1) can be considered as the definition of the commutator of two operators.

Now by considering the aforementioned definitions as well as three axioms, we shall prove the following theorem which is our main result:

**Theorem.** For every Dirac function, the following equation holds:

$$i\hbar \frac{\partial}{\partial t} \psi(\xi t) \Big\rangle = H \psi(\xi t) \rangle,$$

where H denotes the total energy (Hamiltonian) operator of a given dynamical system, and $t$ stands for time (i.e., $\xi t$ depicts that the Dirac function is generally both a function of $\xi$ and $t$).

## 3. Proof of the theorem

Consider a particular state of motion throughout the time during which the system is left undisturbed. We shall have the state at any time, $t$, corresponding to a certain ket depending on $t$ which is denoted by $|t\rangle$. Since we have considered an undisturbed dynamical system, the causality among the states holds and the requirement that the state at one time determines the state at another time, means that, for instance, $|Pt_0\rangle$ (denoted a state at time $t_0$) determines $|Pt\rangle$ (denoted a state at time $t$). Axiom 3 implies that $|Pt\rangle$'s are linear functions of the $|Pt_0\rangle$'s and accordingly, each $|Pt\rangle$ is the result of some linear operator applied to $|Pt_0\rangle$. In other words:

$$|Pt\rangle = T|Pt_0\rangle, \qquad (11)$$

where T is a linear operator independent of $P$ and depends only on $t$ (or $t_0$).

It is straightforward that the definition of bras in this discussion is possible only by making a conjugate imaginary of corresponding kets. Accordingly consider a bra $\langle Qt_0|$. Then the scalar product of this bra with a ket $|Pt_0\rangle$ would be a number, $c$, that is,

$$\langle Qt_0|Pt_0\rangle = c. \tag{12}$$

Axiom 3 states that the superposition relationships between states under displacement remain invariant, therefore

$$\langle Qt|Pt\rangle = c = \langle Qt_0|Pt_0\rangle. \tag{13}$$

**Lemma 2.** The linear operator T is a unitary operator.

*Proof.*

By making conjugate imaginary of (11) and substituting $Q$ instead of $P$, and then finding the scalar product of that with the ket $|Pt\rangle$, one with the help of (13) finds

$$\langle Qt|Pt\rangle = \langle Qt_0|\overline{T}T|Pt_0\rangle = \langle Qt_0|Pt_0\rangle, \tag{14}$$

where $\overline{T}$ denotes the complex conjugate of T. Since (14) holds for any general $P$ and $Q$, one can conclude

$$\overline{T}T = \hat{1}. \tag{15}$$

In addition, by making conjugate imaginary of (14), one leads to

$$\langle Pt|Qt\rangle = \langle Pt_0|T\overline{T}|Qt_0\rangle = \langle Pt_0|Qt_0\rangle. \tag{16}$$

Since (16) holds for any general $P$ and $Q$, thus

$$T\overline{T} = \hat{1}. \tag{17}$$

Equation (15) together with (17) implies that T is a unitary operator (Q.E.D).

Now suppose that the effect of an arbitrary real dynamical variable $\mathfrak{I}$ on the state $|Pt_0\rangle$ is another state denoted by $|Rt_0\rangle$, that is,

$$\mathfrak{I}|Pt_0\rangle = |Rt_0\rangle. \tag{18}$$

If $\mathfrak{I}_t$ denotes the corresponding time-displaced dynamical variable, then based on the Axiom 3 one would be able to write:

$$\Im_t |Pt\rangle = |Rt\rangle \tag{19}$$

In addition, with the help of Eq. (11) and Lemma 2 one can conclude that

$$|Pt_0\rangle = T^{-1}|Pt\rangle. \tag{20}$$

Using Lemma 2, and Eqs. (19), (11), (18) and then (20) one can write:

$$\Im_t |Pt\rangle = |Rt\rangle = T|Rt_0\rangle = T\Im|Pt_0\rangle = T\Im T^{-1}|Pt\rangle \tag{21}$$

Since (21) holds for an arbitrary ket $|Pt\rangle$, hence

$$\Im_t = T\Im T^{-1}. \tag{22}$$

Now we pass to the infinitesimal case by making $t \to t_0$ and assume from the physical continuity as well as with the help of (11) that the following limits exist:

$$\lim_{t \to t_0} \frac{|Pt\rangle - |Pt_0\rangle}{t - t_0} = \lim_{t \to t_0} \frac{T - \hat{1}}{t - t_0} |Pt_0\rangle \tag{23}$$

We define the second limit of (23) as the *time-displacement operator* which is denoted by $d_t$. By defining $t - t_0 \equiv \delta t$, the definition of the time-displacement operator in symbols would be as follows:

$$d_t \equiv \lim_{\delta t \to 0} \frac{T - \hat{1}}{\delta t} \tag{24}$$

**Lemma 3.** The time-displacement operator $d_t$ is a pure imaginary operator.

*Proof.*

For very small (infinitesimal) $\delta t$, (24) can be written as

$$T = \hat{1} + \delta t \, d_t. \tag{25}$$

So with the help of (25), (15), and ignoring the term $(\delta t)^2$, one can simply find

$$(\hat{1} + \delta t \, \bar{d}_t)(\hat{1} + \delta t \, d_t) = \hat{1} \Rightarrow \delta t (\bar{d}_t + d_t) = 0, \tag{26}$$

where $\bar{d}_t$ denotes the complex conjugate of $d_t$. Equation (26) demonstrates that the time-displacement operator $d_t$ is a pure imaginary operator (Q.E.D).

**Remark 5.** It is possible to consider an arbitrary numerical factor $e^{i\theta}$, with $\theta$ real, which one can multiply into T, and it must be made tending to unity as $\delta t$ tends to zero. So it introduces an *arbitrariness* [24] in $d_t$, namely (24) can be rewritten as follows:

$$\lim_{\delta t \to 0} \frac{T e^{i\theta} - \hat{1}}{\delta t} = \lim_{\delta t \to 0} \frac{T - \hat{1} + i\theta}{\delta t} = d_t + i\alpha_t, \tag{27}$$

where

$$\alpha_t \equiv \lim_{\delta t \to 0} \left( \frac{\theta}{\delta t} \right). \tag{28}$$

Moreover, with the help of (25), (22), Lemma 3 and ignoring the term $(\delta t)^2$, one can find

$$\Im_t = (1 + \delta t \, d_t) \Im (1 - \delta t \, d_t) = \Im + \delta t (d_t \Im - \Im d_t). \tag{29}$$

Now let us define the real dynamical variable $\Im$ to be the total energy (Hamiltonian) of a dynamical system in the time *t*, which is denoted by H(*t*). Due to the Axiom 2, such a definition is possible and reasonable, and due to the discussions in the Appendix, the concept of "Hamiltonian" is realizable in the framework. Note that by writing the Hamiltonian operator as H(*t*), or for more brevity H, we do not mean that it is only a function of time. Rather, it can be also a function of canonical coordinates and linear momenta [60], or even may not be an explicit function of time. So, one should read the notation as a general function of canonical coordinates, linear momenta and time (which is assumed to be at least differentiable with respect to time in one order) – unless we introduce the opposite assumption. Accordingly after taking an infinitesimal time $\delta t$, the time-displaced dynamical variable $\Im_t$ can be considered as the Hamiltonian of the system after taking $\delta t$, that is, H(*t* + $\delta t$). Therefore with the help of (29) one obtains

$$H(t + \delta t) = H(t) + \delta t (d_t H(t) - H(t) d_t). \tag{30}$$

Since $\delta t$ is very small (infinitesimal), (30) can be rewritten as follows:

$$\lim_{\delta t \to 0} \frac{H(t + \delta t) - H(t)}{\delta t} = d_t H(t) - H(t) d_t \tag{31}$$

The left-hand side of (31) is evidently the differentiation of the Hamiltonian with respect to time. In consequence, (31) can be written as

$$\frac{d H(t)}{dt} = d_t H(t) - H(t) d_t, \tag{32}$$

describing the general dynamics of the Hamiltonian in terms of the time-displacement operator.

Now once again, consider (22) and substitute the Hamiltonian of the system instead of the real dynamical variable $\Im$. In order to continue the proof, we shall consider two distinct cases as follows:

**Case I.** Suppose that the original Hamiltonian, H, is not an explicit function of time. Therefore, after an infinitesimal time-displacement one can write

$$H(t + \delta t) = H(t). \tag{33}$$

Hence (22) leads one to

$$T H(t) T^{-1} = H(t), \tag{34}$$

and with the help of Lemma 2, one obtains

$$T H(t) = H(t) T. \tag{35}$$

Differentiating both sides of (35) with respect to time, yields

$$\frac{dT}{dt} H + T \frac{dH}{dt} = \frac{dH}{dt} T + H \frac{dT}{dt},$$

$$\left( \frac{dT}{dt} H - H \frac{dT}{dt} \right) = \left( \frac{dH}{dt} T - T \frac{dH}{dt} \right),$$

$$\left[ \frac{dT}{dt}, H \right] = \left[ \frac{dH}{dt}, T \right]. \tag{36}$$

Then by the help of Lemma 1 (or Remark 4) the result is as follows:

$$\left\{ \frac{dT}{dt}, H \right\} = \left\{ \frac{dH}{dt}, T \right\} \tag{37}$$

Using (37) and with the help of (25) one finds

$$\left\{ \frac{d(d_t)}{dt}, H \right\} = \left\{ \frac{dH}{dt}, d_t \right\}. \tag{38}$$

In order that (38) to be fulfilled,

$$H = a\, d_t, \tag{39}$$

where *a* is a constant number that we shall find it. Note that we already proved that the time-displacement operator is a pure imaginary operator (Lemma 3), and also we mentioned that the total energy of the system is considered as an observable (Axiom 2). Consequently, (39) leads one to conclude that the constant *a* must be a pure imaginary number. In addition, the left-hand side of (39) has the dimension of energy, i.e. $m^2 kg/(sec)^2$ in the international system of units, however the right-hand side is of dimension $(sec)^{-1}$. Since we have considered a general Hamiltonian (which is not an explicit function of time in the current Case) as well as a general time-displacement operator, we will need a universal constant with the physical dimension of $m^2 kg/sec$. The only universal constant with that dimension – which is known so far – is the Planck's constant. Accordingly, $a = i\hbar$, and (39) gives

$$H = i\hbar\, d_t. \tag{40}$$

**Remark 6.** Equation (40) implies that the difference, $H - i\hbar\, d_t$, commutes with all the dynamical variables in a given representation. It may be argued that (40) does not necessarily hold, as its right-hand side can be added with a constant operator, say $b\hat{1}$, where *b* is a real number. It should be noted that with the help of Remark 5, the difference can be made zero by a suitable choice of the arbitrariness [i.e. Eq. (28)] so that (40) remains valid.

We already discussed that from the physical continuity, the limits (23) should exist. The left-hand side of (23) is evidently the differentiation of $|Pt_0\rangle$ with respect to time $t_0$. Therefore (23) with the help of (24) can be rewritten as follows:

$$\frac{d|Pt_0\rangle}{dt_0} = \lim_{t \to t_0} \frac{|Pt\rangle - |Pt_0\rangle}{t - t_0} = \lim_{t \to t_0} \frac{T - \hat{1}}{t - t_0}|Pt_0\rangle = d_t|Pt_0\rangle \tag{41}$$

By Multiplying both sides of (41) by $i\hbar$ and using (40), one concludes for general time, *t*, that

$$i\hbar \frac{d|Pt\rangle}{dt} = H|Pt\rangle. \tag{42}$$

Now introduce a representation with a complete set of commuting observables $\xi$ diagonal (see Remark 1). By putting $\langle \xi'|Pt\rangle$ equal to $\psi(\xi't)$ (Definition 2), the ket $|Pt\rangle$ can be written as $|\psi(\xi t)\rangle$ using the standard ket notation. Consequently (42) can be rewritten as:

$$i\hbar \frac{\partial}{\partial t}|\psi(\xi t)\rangle = H|\psi(\xi t)\rangle, \tag{43}$$

which proves the Theorem for the Case I.

**Case II.** Suppose that the original Hamiltonian, H, is an explicit function of time, $t$, that is, $H(q,p,t)$, where $q$ and $p$ denote canonical coordinates and linear momenta, respectively. In this case, one can formally construct a new time-independent Hamiltonian, $\tilde{H}(\tilde{q},\tilde{p})$, where $\tilde{q}$ and $\tilde{p}$ denote the new coordinates and new linear momenta, respectively, on an extended phase space according to the following embedding [51]:

$$\tilde{q}_r = \begin{cases} q_r & 1 \le r \le M \\ t & r = 0 \end{cases} \tag{44-a}$$

$$\tilde{p}_r = \begin{cases} p_r & 1 \le r \le M \\ -E & r = 0 \end{cases} \tag{44-b}$$

$$\tilde{H}(\tilde{q},\tilde{p}) = H(q,p,t) - E, \tag{44-c}$$

where $E$ and $t$ are considered to be two new conjugate variables, $M$ is an integer number and can depict the number of degrees of freedom, and the integer index $r$ runs from zero to $M$.

The associated new Hamilton equations in classical mechanics are as follows:

$$\dot{\tilde{q}}_r = \frac{\partial \tilde{H}(\tilde{q},\tilde{p})}{\partial \tilde{p}_r}, \tag{45-a}$$

$$\dot{\tilde{p}}_r = -\frac{\partial \tilde{H}(\tilde{q},\tilde{p})}{\partial \tilde{q}_r}. \tag{45-b}$$

Evidently for $1 \le r \le M$, both new and old equations of motion are identical. For the new conjugate variables, $t = \tilde{q}_0$ and $-E = \tilde{p}_0$. One can with the help of (44)'s write:

$$\dot{\tilde{q}}_0 = \frac{\partial \tilde{H}(\tilde{q},\tilde{p})}{\partial \tilde{p}_0} = 1 \tag{46-a}$$

$$\dot{\tilde{p}}_0 = -\frac{\partial \tilde{H}(\tilde{q},\tilde{p})}{\partial \tilde{q}_0} = -\frac{\partial H(q,p,t)}{\partial t} \tag{46-b}$$

Using (44)'s it is seen that (46-a) is identical with $\dot{t} = \partial[H(q,p,t) - E]/\partial(-E) = 1$, and (46-b) is identical with $\dot{E} = \partial H(q,p,t)/\partial t$. In addition, it shows that if the original Hamiltonian were time-independent, then the variable $t$ would be cyclic (which completely makes sense).

Note that, it is now possible to incorporate the new coordinates $\tilde{q}$'s and new linear momenta $\tilde{p}$'s as a new variable, call it $\zeta$, which is any function of them and does not contain the time $t$ explicitly. So one can write

$$\frac{d\zeta}{dt} = \sum_r \left( \frac{\partial \zeta}{\partial \tilde{q}_r} \frac{d\tilde{q}_r}{dt} + \frac{\partial \zeta}{\partial \tilde{p}_r} \frac{d\tilde{p}_r}{dt} \right), \tag{47}$$

and with the help of (45)'s it can be rewritten as

$$\frac{d\zeta}{dt} = \sum_r \left( \frac{\partial \zeta}{\partial \tilde{q}_r} \frac{\partial \tilde{H}}{\partial \tilde{p}_r} - \frac{\partial \zeta}{\partial \tilde{p}_r} \frac{\partial \tilde{H}}{\partial \tilde{q}_r} \right) = \{\zeta, \tilde{H}\}. \tag{48}$$

To use the result in the framework, (48) can with the help of Lemma 1 be mapped into the commutator and the variables shall be considered as operators. Thus in case of time-dependent Hamiltonian, one is able to follow the aforementioned discussions to get one Hamiltonian which is implicitly a function of time. With the help of this result and also by using Remark 1 – which can lead one to define a new representation in which a complete set of commuting observables are diagonal – the Hamiltonian is not an explicit function of time anymore. This allows the following conclusion:

$$\tilde{H}(t+\delta t) = \tilde{H}(t), \tag{49}$$

by noting that, now the notation $\tilde{H}(t)$ or $\tilde{H}(t+\delta t)$ shows that the Hamiltonian does not depend explicitly on time. The proof for this second case follows identically from equations (34) to (43) by taking into account that H is replaced by $\tilde{H}$. By considering the arguments of both cases I and II, we infer that the proof of the Theorem is complete (Q.E.D).

**Remark 7** (Interpretation of the Dirac function). Equation (43) is the so-called non-relativistic *time-dependent Schrödinger wave equation.* The Eq.(43) also implies that the Dirac function (Definition 2) and the *wave function* of the traditional formalism are identical. Therefore the traditional wave function is defined and explained by Definition 2 [24]. By the way, the question of how such a Dirac function or equivalently, wave function (or more generally the whole theory) can be interpreted physically, is not going to be discussed here. Historically, Born and Jordan considered quantum theory as being essentially a statistical theory [18,38], an approach that Einstein could never accept [18]. Hereby the statistical element was only a result of the performed experiments interpreted in the sense of Heisenberg and Dirac [10,11,38]. In accordance with the above proof, one is free to employ the statistical interpretation. One can consider the numbers, which form the representative of a normalized ket or bra (see Definition 2), as probability amplitudes (see, for instance, sections 9-12 and 20 of [24], or [30] for justifications). Furthermore, one is free to accept the mentioned physical interpretations of the whole theory. Nevertheless if someone says that a given approach needs a new axiom in our framework, the reply will be: answering this question is essentially a fundamental problem on another level of discussion about quantum mechanics which is not the main aim of this paper. Moreover, the number of axioms in our framework will remain less than the number of postulates in the usual Schrödinger picture. This is because in the usual Schrödinger picture, the Schrödinger equation and also the ordinarily given interpretations are based on postulates.

## 4. Technical discussion

As noted in Section 1, a great number of works done so far to derive the Schrödinger equation are not satisfactory as of yet. Those attempts can be divided into three general groups. In first two (discussed in Section 1), one either finds no precise formalisms, or they simply deal with classical wave theory mislead by the word 'wave function' or 'wave equation' – which fact already caused both Dirac and Schwinger to express their dissatisfaction with them. The third group is the one addressed here. So the question here is why our approach is distinct, and what the advantage of our viewpoint – beyond other precise works done so far – is.

The main aim of our work was to derive the time-dependent Schrödinger equation from just three axioms, that is, to reduce one of the existing four axioms of the Schrödinger picture to a theorem. We could also show what the precise physical meaning of the term 'Hamiltonian' in our framework is (see Appendix). Both the axioms and the method of introducing the Hamiltonian in the framework were made transparent *physically*.

The number of works in the third group which have tried to present a precise formalism can be divided into four categories:

1) In spite of starting out with a precise mathematical physics framework, the Schrödinger equation here is exemplified by the traditional wave mechanics (see, for instance, Ref. [25]). We showed in Section 1 why such a point of view is not satisfactory.

2) Many axioms are introduced which have no direct physical meaning (e.g. [15,26,44,48,54]). Recently, Kapustin [55] has commented on such works, although the remedies offered by him are still marred by some non-palpable physical axioms and the exact physical meaning of his Hamiltonian and the role of Planck's constant remain vague. Unfortunately, the number of axioms is increased rather than reduced, so that the strength and reliability of the theory suffers.

3) The Schrödinger equation is only a postulate (for instance, Refs. [31,49]), or else Eq. (40) is postulated as implicitly done Ref. [20]. In contrast our aim was to show that the Schrödinger equation is not a postulate but can be *derived* from the other three axioms of the Schrödinger picture.

4) In the final fourth category, the Schrödinger equation with its exact coefficients (including Planck's constant) is not successfully derived. Moreover, a large number of works in this area (e.g. [26,29,39,42,52]) are based on Stone's theorem [12,52] in which a self-adjoint operator in

Hilbert space, say H, appears and is called the Hamiltonian. Although the theorem does state that such an operator is unique, it does not give a physical meaning to this operator. In other words, the procedure still needs a proof that the operator appearing in Stone's theorem is the same physical Hamiltonian which can be established by an 'analogy' to classical physics (see Appendix).

Similarly, Glimm and Jaffe [39] have proved a theorem called 'reconstruction of quantum mechanics' by using the Osterwalder-Schrader axioms [32,33] in which a self-adjoint operator, say H, based on the probability measure appears and is called the Hamiltonian. This procedure will only be complete once a uniqueness proof for that operator is presented. Even then it is still open what the physical meaning of such an operator is and whether it is equivalent to the physical Hamiltonian which is discussed in the Appendix.

Amongst the works in this category, Thirring [46] presents a more physical introduction for the Hamiltonian in quantum mechanics. He introduces Hamiltonian by using an analogy to classical mechanics in which Hamiltonian generates the time-evolution. This analogy is implicitly one of his postulates. In contrast to him, we could *derive* such dynamics for our deduced Hamiltonian in Eq. (32). Thirring moreover supports this idea using two added postulates, namely: (i) the dynamics of a closed system can be described quantum-mechanically by an equation of the form $(d/dt)f = af$, where $a$ is a time-independent operator; and (ii) the algebra of observables evolves in time according to $b(t) = \exp(i\mathrm{H}t)b\exp(-i\mathrm{H}t)$, where H denotes the Hamiltonian [46]. In contrast, we assumed *none* of those axioms in our own framework to show a physical meaning to the Hamiltonian.

It should be noted that the main problem that we posed to solve, was not showing a new way to introducing the Hamiltonian in quantum mechanics (nonetheless two precise and physical approaches are introduced in Appendix). Rather, the key problem was to prove that the time-dependent Schrödinger equation is not a postulate, because it can be *derived* from the three remaining axioms of the Schrödinger picture. This problem was addressed in *none* of the mentioned works.

## 5. Concluding remarks

It has been shown that the time-dependent Schrödinger equation can be deduced from three axioms: (i) spatial coordinates (positions) are all observables and have continuous ranges of eigenvalues; (ii) the total energy of a dynamical system (its Hamiltonian) is always an observable; and (iii) the superposition relationship between states remains invariant under displacements. Therefore, regardless of the physical interpretation of quantum mechanics, the number of axioms needed in the theory is reduced from four to three. For it was possible to eliminate the time-dependent Schrödinger equation as a postulate and make it a *theorem*. As a direct consequence, validity of the time-dependent Schrödinger equation can henceforth no longer be falsified by a counter-experiment as was so far possible in principle. Thus, the whole theory has become stronger. Perhaps it will sometime be possible to disprove some of the remaining axioms used above, or else find a novel phenomenon which implies that the theorem needs radical modification. However, as discussed in Sections 1 and 4, the proof presented here is immune to the problems affecting previous attempts in the same direction since it arose in an explicit framework with clearly defined key concept. In particular, the method by which the equation was derived does not depend on the wave approach. So, one can delineate the following steps: (a) definition of the framework; (b) proof of the non-relativistic time-dependent Schrödinger equation as a theorem; (c) freedom to add other properties and move on freely to the other pictures and come back using known techniques.

It is of great interest to me that recently a collection of open problems on the foundations of quantum physics – called the "Oxford Questions" – have been published [58]. Here problem (4.a) is as follows: 'What insights are to be gained from category-theoretic, informational, geometric and operational approaches to formulating quantum theory?' This question can be considered as a special case of the sixth Hilbert problem [47] which reads: "The investigations on the foundations of geometry suggest the problem: to treat in the same manner, by means of axioms, those physical sciences in which mathematics plays an important part, in the first rank are the theory of probabilities and mechanics." Although a great number of works in this area have been done, the recent "Oxford Questions" accentuate that the task has not yet been satisfactorily solved in the case of quantum mechanics and that therefore the sixth Hilbert problem for quantum mechanics is still considered open. The effort of the present paper to employ three non-artificial (rather palpable and physical) axioms in order to successfully derive the time-dependent Schrödinger equation, which leads to the elimination of one of the elegant axioms of quantum mechanics and allows one to move from the Schrödinger picture to other pictures by known techniques, can perhaps be considered as a step to answering part of problem (4.a) posed in the Oxford Questions.

## Acknowledgements

I thank Nils Schopohl for very helpful discussions and suggestions.

## Appendix

### A.1. What is the subtle point?

Remark 2 refers to a subtle point of Axiom 2: The proposition that "the total energy (Hamiltonian) of a dynamical system is always an observable," does not imply that the concept of 'Hamiltonian,' itself, is clear in the framework. To define the concept of Hamiltonian, we assume that the concept of a 'particle' or an 'elementary particle' is a primitive concept or term.

Nevertheless it can be defined precisely (e.g., see Refs. [21,35,37]). But in our framework it is not necessary to go to a deeper level. For atomic structure discussions, the elementary particles are electrons and nuclei, for nuclear physics discussions, they are protons and neutrons, etc. [43]. In analogy, the concept of 'position' is also level dependent. In order to employ the concept of Hamiltonian in the present framework, we have at least two different ways: one derived from the Dirac formalism [24], and the other by Schwinger [43]. Both have a close analogy to classical mechanics and that 'analogy' or 'proportionality' has played a pivotal role in the history of quantum mechanics (e.g. [2,11,14,16-18,27,53]).

### A.2. Dirac's approach

The Dirac formalism uses the concept of Hamiltonian in quantum mechanics based on the Axiom 1 and Lemma 1. To understand it better, one can start out with the concept of 'linear momentum' in analogy with classical mechanics as an essential characteristic of a particle. Using the following equalities from classical mechanics,

$$\{q_r, q_s\} = \{p_r, p_s\} = 0,$$
$$\{q_r, p_s\} = \delta_{rs}, \tag{A1}$$

where $r$ and $s$ are two integer numbers counting from 1 to $n$ (the number of degrees of freedom of a given dynamical system), then one can have the following corresponding quantum versions with the help of Lemma 1:

$$[q_r, q_s] = [p_r, p_s] = 0,$$
$$[q_r, p_s] = i\hbar \{q_r, p_s\} = i\hbar \delta_{rs}. \tag{A2}$$

Using Definition 2 and by employing the standard ket notation, an arbitrary ket can be written as $\psi(q_1...q_n)\rangle = \psi\rangle$. Furthermore, one can introduce $n$ linear operators, namely, $\partial/\partial q_r$ $(r = 1,...,n)$, which can operate on an arbitrary ket based on the following relations:

$$\frac{\partial}{\partial q_r} \psi \Big\rangle = \frac{\partial \psi}{\partial q_r} \Big\rangle,$$
$$\frac{\partial}{\partial q_r} \Big\rangle = 0. \tag{A3}$$

Next, by taking representatives (see Remark 1), using (A3) and the general theory of linear operators one can conclude [24]:

$$\forall \psi\rangle: \quad \frac{\partial}{\partial q_r} q_s \psi \Big\rangle = \frac{\partial q_s \psi}{\partial q_r} \Big\rangle = q_s \frac{\partial}{\partial q_r} \psi \Big\rangle + \psi \rangle , \tag{A4}$$

yielding

$$\frac{\partial}{\partial q_r} q_s - q_s \frac{\partial}{\partial q_r} = \left[\frac{\partial}{\partial q_r}, q_s\right] = \delta_{rs} .$$ (A5)

Comparing (A5) with the second line of (A2), it would be possible to take

$$p_r = -i\hbar \frac{\partial}{\partial q_r} ,$$ (A6)

by fixing the representation except for an arbitrary constant phase factor. Then using (A6), one is able to consider a number of free elementary particles and establish the total energy of them by ascribing mass to each of them in analogy to classical mechanics. Then as the next step, add an arbitrary potential describing the interaction energy. So the Hamiltonian can be modified by adding that potential to the former total energy, and therefore the concept of Hamiltonian can be understood in the present framework.

### A.3. Schwinger's approach

The second approach is the Schwinger's, stating that it is possible to suppose that an elementary particle can be described by position, linear momentum, mass, and spin. Then it would be feasible to find the total mass of a number of elementary particles, and accordingly ascribe a center of mass to them. Afterwards one can form the total linear and angular momenta. In Schwinger's approach, a new vector quantity, that may be called 'booster,' is defined and does not have a classical analogous. In other words, if $\mathbf{r}_k$ and $\mathbf{p}_k$ denote the vector of position and linear momentum of the $k$-th elementary particle, respectively, then the new vector quantity booster (denoted by $\mathbf{N}$) can be defined as follows:

$$\mathbf{N} \equiv \sum_k \left(\mathbf{p}_k t - m_k \mathbf{r}_k\right) ,$$ (A7)

which can then be written as $\mathbf{P}t - M\mathbf{R}$, where $\mathbf{P}$ is the vector of total linear momentum, $t$ stands for time, $M$ denotes the sum of all masses, and $\mathbf{R}$ is the center of mass. Furthermore, by writing the total vector of angular momentum, one can ignore the terms showing internal variables of the free elementary particles, and therefore, the term corresponding to the total internal angular momentum is calculated. Then the next step becomes identical to Dirac's approach, namely, first it is possible to define the Hamiltonian when the constituents are isolated from each other, and then add the potential interaction energy, which is a scalar function of the internal variables, spin, etc., so that the concept of Hamiltonian is well specified in the present framework.

### References:

[1] Wentzel, G. 1924 Zur Quantenoptik. *ZS f. Physik*. **22**, 193-199.

[59] In Ref. [4] where Dirac derived Eq. (1) for the first time, he found the coefficient $\Omega$ by an 'assumption,' that is, he mentioned there: "we make the fundamental assumption that the difference between the Heisenberg products of two quantum quantities is equal to $ih/2\pi$ times their Poisson bracket expression;" however, he did not use the word 'assumption' in Ref. [24] anymore. Nevertheless, here we present a discussion to find the coefficient $\Omega$.

[60] To understand what the meaning of 'linear momentum' in the framework is, see the subsection A.2 of Appendix.